\documentclass{elsart}
\usepackage{graphicx} 
\usepackage{amsmath}
\usepackage{rotating}

\begin{document}
\begin{frontmatter}

\title{Configuration Interaction calculations of positron binding to Be($^3$P$^o$)}
\author{M.W.J. Bromley}
\address{Department of Physics, San Diego State University, San Diego CA 92182, USA}
  \ead{mbromley@physics.sdsu.edu}

\author{J. Mitroy}
\address{Faculty of Technology, Charles Darwin University, Darwin NT 0909, Australia}
  \ead{jxm107@rsphysse.anu.edu.au}


\begin{abstract}
The Configuration Interaction method is applied to investigate 
the possibility of positron binding to the metastable beryllium
($1s^22s2p$ $^3$P$^o$) state.  The largest calculation obtained an
estimated energy that was unstable by 0.00014 Hartree with respect
to the Ps + Be$^+$(2s) lowest dissociation channel.  It is 
likely that positron binding to parent states with non-zero 
angular momentum is inhibited by centrifugal barriers.   
\end{abstract}

\end{frontmatter}

\section{Introduction}

In the last several years there has been substantial progress studying
the physics of positron binding to atoms.  As late as 1997,
there was no rigorous evidence for existence of positron-atom
bound states \cite{schrader98}, although the possibility of 
positron binding had been often invoked  
to explain the existence of very large annihilation
rates for positrons annihilating in gases 
\cite{paul63a,khare64,goldanskii64a,mcnutt71a,surko88}.   Towards the end 
of 1997, positron binding to atomic lithium was established 
in two independent calculations \cite{ryzhikh97b,strasburger98}.  
Following these initial studies, positron binding to a further
10 elements of the periodic table was demonstrated \cite{mitroy02b}.
One of the features of these 
positronic atoms is that all the atoms binding a positron have 
ionization potentials close to 6.80 eV (the Ps binding energy) 
and the binding energies are generally larger for the atoms with
their ionization energies closest to 6.80 eV.     

A related question is whether positrons can form bound states 
with excited electronic states.  There is one excited atomic 
state known to bind a positron, namely the metastable 
He($^3$S$^e$) state \cite{ryzhikh98d,mitroy05e}.  
The stability of $e^+$He($^3$S$^e$) state was established 
with the Stochastic Variational Method (SVM) 
\cite{varga95,suzuki98a,ryzhikh97b} following an initial  
inconclusive study by Drachman {\em et al} \cite{drachman76}.  

The answer to this question is relevant to positron-atom collision physics 
\cite{surko05a}, since such states may manifest themselves as Feshbach
resonances in the elastic or excitation cross sections. 
There has been a suggestion that the rich resonant structures 
prominent in electron-atom scattering \cite{buckman94} are largely absent
from the positron-atom spectrum \cite{sullivan01b,mitroy02b,surko05a}.  

This work investigates the possibility of positron binding to
the low-lying triplet odd state of neutral beryllium, 
i.e. the Be($1s^22s2p$ $^3$P$^o$) system.  This state has been identified
as one of the most likely excited states to bind a positron 
\cite{schrader01a,mitroy02b}.  This conjecture has largely 
been based upon the fact that its ionization potential (IP) 
of 6.598 eV \cite{bashkin75} is quite close to the Ps
binding energy. This IP is between that of magnesium and calcium, both
of which bind a positron with a binding energy of 
$\approx$ 0.4 eV \cite{mitroy02b,bromley02b,bromley06a}.
Furthermore, the dipole polarisability of Be($^3$P$^o$) is
$\alpha_d=39.02$ $a_0^3$ \cite{mitroy04b}, greater than that of 
ground state Be ($\alpha_d=37.7$ $a_0^3$ \cite{mitroy03f}).

This metastable state warrants investigation for another reason;   
the parent atom has an angular momentum greater than zero with 
the loosely bound $2p$ electron carrying the angular momentum.
The quantitative impact that the angular momenta has on the ability to 
bind a positron is largely unknown.  Angular momentum coupling 
considerations suggests that the existence of a repulsive 
centrifugal barrier will act to inhibit positron binding by 
making it less energetically favorable to form a Ps-cluster.  
Either, the electron-positron pair will be in a relative $p$-state
with a smaller energy gain than the Ps($1s$) state, or the
total angular momentum of the Ps cluster relative to the nucleus
will be greater than zero.  

\section{Details of the Calculation}

\begin{figure}[bht]
\centering{  
\includegraphics[width=8.5cm,angle=0]{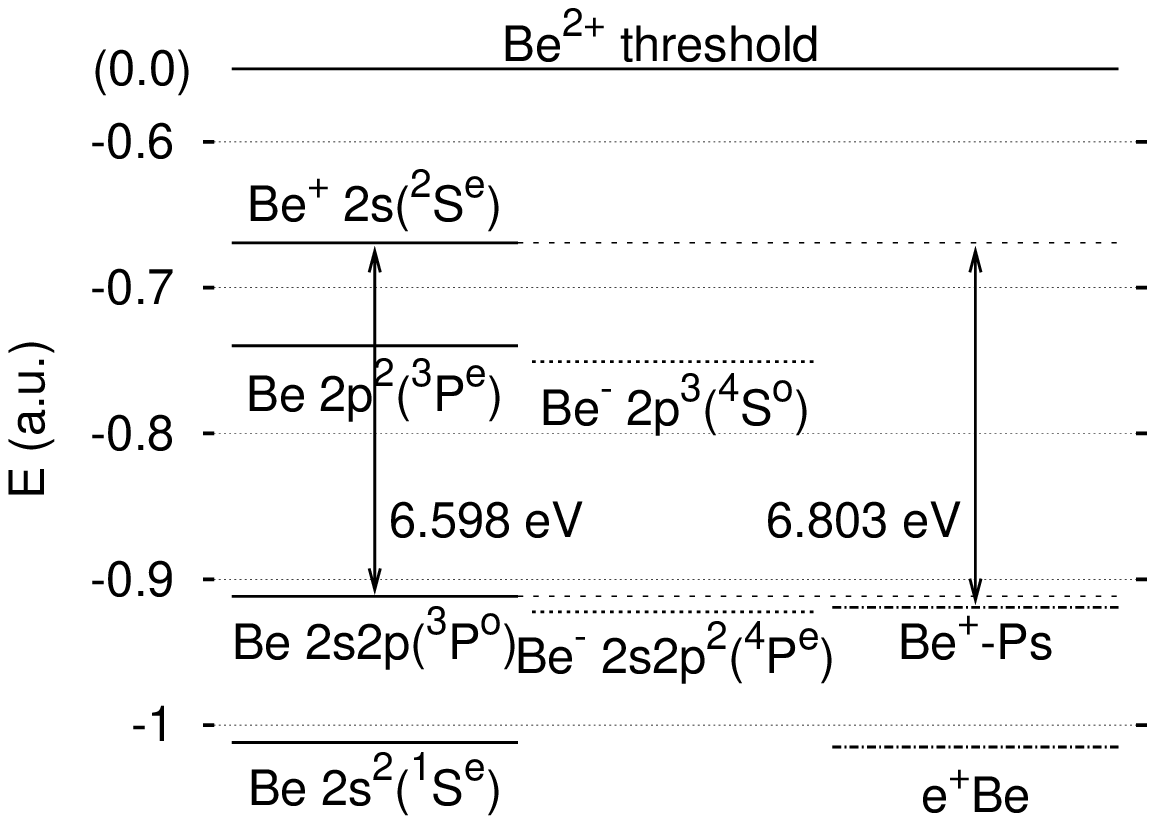}
}
\caption[]{ 
The energy levels of electron and positron binding to neutral beryllium
(in units of Hartree relative to the Be$^{2+}$ threshold).  
The experimental atomic binding energies are taken from the NIST compilation
\cite{nistasd3}, the metastable negative ion experimental binding energies are 
from \cite{andersen97a}, and the theoretical $e^+$Be binding energy is taken
from a frozen-core SVM calculation \cite{mitroy01c}.
The threshold for $e^+$ binding to Be($^3$P$^o$) is denoted as Be$^+$+Ps.
}
\label{fig:beenergies}
\end{figure}

Since the ionization energy of the Be($2s2p$ $^3$P$^o$) system is
just less than 6.80 eV, the threshold for a stable positron complex
is that for the Be$^+$($2s$) +  Ps($1s$) dissociation channel.  This 
energy was $-0.9192086$ Hartree for the present model potential.  An 
energy level diagram showing the ground and low-lying metastable states 
of Be, and some negative ion and positronic ion energies are shown 
in Figure~\ref{fig:beenergies}.     

The CI method as applied to positron-atomic systems with two
valence electrons and a positron has been discussed previously  
\cite{bromley00a,bromley02a,bromley02b}, but a short 
description is worthwhile.
The model Hamiltonian is initially based on a Hartree-Fock (HF) wave
function for the neutral atom ground state.
One- and two-body semi-empirical polarization potentials are added 
to the potential field of the HF core and the parameters of 
the core-polarization potentials defined by reference to 
the spectrum of Be$^+$ \cite{norcross76}.

All calculations were done in the frozen-core 
approximation.  The general effective Hamiltonian for the system 
with $N_e$ valence electrons and a positron was 
\begin{eqnarray}
H  &=&  - \frac{1}{2}\nabla_{0}^2 - \sum_{i=1}^{N_e} \frac {1}{2} \nabla_{i}^2 
    + \sum_{i<j}^{N_e} \frac{1}{r_{ij}}
    + \sum_{i=1}^{N_e} (V_{dir}({\bf r}_i) + V_{exc}({\bf r}_i) + V_{p1}({\bf r}_i)) \nonumber \\  
   &-& \sum_{i=1}^{N_e} \frac{1}{r_{i0}} - V_{dir}({\bf r}_0) + V_{p1}({\bf r}_0)
    - \sum_{i<j}^{N_e} V_{p2}({\bf r}_i,{\bf r}_j)
    + \sum_{i=1}^{N_e} V_{p2}({\bf r}_i,{\bf r}_0) \; .
\end{eqnarray}
The direct potential ($V_{dir}$) represents the interaction 
with the HF $1s^2$ electron core.
The direct part of the core potential is attractive 
for electrons and repulsive for the positron.  The exchange 
potential  ($V_{exc}$) between the valence electrons and the 
HF core was computed without approximation.

The one-body polarization potential ($V_{p1}$) was a semi-empirical
polarization potential derived from an analysis of the Be$^+$ 
spectrum.  It has the functional form
\begin{equation}
V_{p1}(r)  =  -\frac{\alpha_d g^2(r)}{2 r^4} .
                                    \label{polar1}
\end{equation}
The factor $\alpha_d$ is the static dipole polarizability of
the core and $g^2(r)$ 
is a cutoff function designed to make the polarization
potential finite at the origin.  The same cutoff function has been
adopted for both the positron and electrons.  In this work, 
$g^2(r)$ was defined to be
\begin{equation}
g^2(r) = 1-\exp\bigl(-r^6/\rho^6 \bigr) \ ,
                                    \label{cut1} 
\end{equation}
where $\rho$ is an adjustable cutoff parameter.  
The core dipole polarizability was set to 0.0523 $a_0^3$ while 
$\rho$ was set to 0.95 $a_0$ \cite{bromley02a}.
The two-body polarization potential ($V_{p2}$) is defined as
\begin{equation}
V_{p2}({\bf r}_i,{\bf r}_j) = \frac{\alpha_d} {r_i^3 r_j^3}
({\bf r}_i\cdot{\bf r}_j)g(r_i)g(r_j)\ . 
                                    \label{polar2}    
\end{equation}
This model potential gives a Be(2s2p $^3$P$^o$) binding energy
of 0.91147 Hartree with respect to the Be$^{2+}$ threshold.
The experimental binding energy with respect to this threshold 
is 0.91888 Hartree \cite{nistasd3}.

The CI basis was constructed by 
letting the two electrons and the positron form all the possible
total angular momentum $L_T = 1$ configurations, with the
two electrons in a spin-triplet state, subject to the selection rules,
namely
\begin{align}
\max(\ell_0,\ell_1,\ell_2) & \le L_{\rm max} \\
\min(\ell_1,\ell_2)& \le L_{\rm int}  \\  
(-1)^{(\ell_0+\ell_1+\ell_2)} & = -1 
\end{align}
In these rules $\ell_0$, $\ell_1$ and $\ell_2$ are respectively 
the orbital angular momenta of the positron and the two electrons.  

Our two-electron-positron calculations with non-zero total
angular momentum were first validated against the previous
$L_T=1$ and $2$ PsH calculations of Tachikawa \cite{tachikawa01a}.
Using their Gaussian-type orbitals we reproduced their reported energy and
annihilation rates.  Note that the PsH states with $L_T=1$ and $2$ are
unbound \cite{bressanini98c,tachikawa01a,saito03a}.

For the $e^+$Be($^3$P$^o$) calculations, the Hamiltonian was
diagonalized in a CI basis constructed from a very large number
of single particle orbitals, including orbitals up to $\ell = 12$.
There was a minimum of $12$ radial basis functions for each $\ell$.
The largest calculation was performed with $L_{\rm max} = 12$ and
$L_{\rm int} = 3$ and gave a CI basis dimension of 498750.  The resulting 
Hamiltonian matrix was diagonalized with the Davidson algorithm \cite{stathopolous94a},
and a total of 1376 iterations were required for the largest calculation.

\section{Results}

Various $e^+$Be($^3$P$^o$) expectation values, such as energy and mean distance
of the electron and positron from the nucleus are given in Table 
\ref{tab:belmax}.  The $2\gamma$ annihilation rate \cite{mitroy02b} for
annihilation with the core and valence electrons are denoted $\Gamma_c$ and 
$\Gamma_v$ respectively.
The calculations shown in Table \ref{tab:belmax} have a minimum of $12$ radial
basis functions for each $\ell$ with $L_{\rm max} = 12$ and $L_{\rm int} = 3$.
The largest calculation remains unbound by $-0.0021904$ Hartree.

To demonstrate that $L_{\rm int} = 3$ is sufficient for this system,
also shown in Table \ref{tab:belmax} is the result of a
$L_{\rm max} = 12$ and $L_{\rm int} = 2$ calculation.
The $L_{\rm max}$ parameter needs to be large since it determines the
extent to which electron-positron correlations are incorporated into
the wavefunction.  However, $L_{\rm int}$ is largely concerned with 
electron-electron correlations and, for example, setting $L_{\rm int} = 3$
for the PsH and $e^+$Be ground states recovered 99.4$\%$ and 98.5$\%$  
respectively of the Ps and positron binding energies for a given 
$L_{\rm max}$ \cite{bromley02a}.

The main problem afflicting CI calculations of positron-atom interactions 
is the slow convergence of the expectation values with $L_{\rm max}$ \cite{mitroy99c,dzuba99,mitroy02b,mitroy06a}.
One way to determine the $L_{\rm max} \rightarrow \infty$ expectation values is 
to assume that the successive increments, $\Delta X_{L}$, to any expectation 
value $\langle X \rangle $ scale as $1/L^p$ as $L$ increases \cite{mitroy99c}.
However, arguments based on 2nd-order perturbation theory suggests that the 
energy increments scale asymptotically as 
$1/(L+\frac12)^{p_E} \approx 1/(L+\frac12)^{4}$ \cite{schwartz62a,gribakin02a}.
While existing CI calculations are consistent with this idea 
\cite{bromley02a,bromley03a,mitroy06a,bromley06c}, the actual size of the 
exponent, $p_E$ is significantly smaller than 4 at $L_{\rm max} \approx 10$.
This can be seen in Table \ref{tab:belmax}, where the power law factor $p_E$ 
relating successive $E(L_{\rm max}\!-\!2)$, $E(L_{\rm max}\!-\!1)$, and 
$E(L_{\rm max})$ are tabulated.
Applying a simple power law extrapolation with $p_E = 2.078$ leads to
a prediction of binding ($E=-0.9207021$), but not much credence should be
placed in this since fixing $p_E$ at its $L_{\rm max} = 12$ value of 2.078 will
result in an extrapolation that will overestimate the contribution
of the orbitals with $L > L_{\rm max}$ \cite{bromley02d,mitroy06a}.  

An improved method, borrowing from existing work in atomic structure 
physics \cite{carroll79a,hill85a,bromley06a}, is to estimate
the $L_{\rm max} \rightarrow \infty$ limit for the energy by assuming that 
\begin{equation}
\Delta E_L \approx \frac {A_E}{(L+{\scriptstyle \frac{1}{2}})^4} 
    + \frac {B_E}{(L+{\scriptstyle \frac{1}{2}})^5} 
    + \frac {C_E}{(L+{\scriptstyle \frac{1}{2}})^6}  \; .
\label{extrap1}
\end{equation}
The factors $A_E$, $B_E$ and $C_E$ are determined from four calculations at 
successively larger values of $L_{\rm max}$.  Applying eq.~(\ref{extrap1}) to 
the data in Table \ref{tab:belmax} at $L_{\rm max}=12$ one finds
$A_E = -34.601$, $B_E = 477.71$ and $C_E=-1829.3$, with
$E=-0.919070$ Hartree as our present best energy estimate.
This is $0.0001390$ Hartree higher the threshold for binding at $-0.9192086$ Hartree.

Usage of an inverse power series restricted to the first two terms
of eq.~(\ref{extrap1}) results in $A_E=-21.875$ and  $B_E = 172.29$, and an
$e^+$Be($^3$P$^o$) energy of $E=-0.9188267$, which is also unbound.
As has been seen in other CI calculations of positron-atom systems,
even when the three-term and the two-term coefficents are significantly different,
their extrapolated energies can still lie relatively close together \cite{mitroy06a}.

\begin{figure}[bht]
\centering{
\vspace{0.2cm}
\includegraphics[width=9.0cm,angle=0]{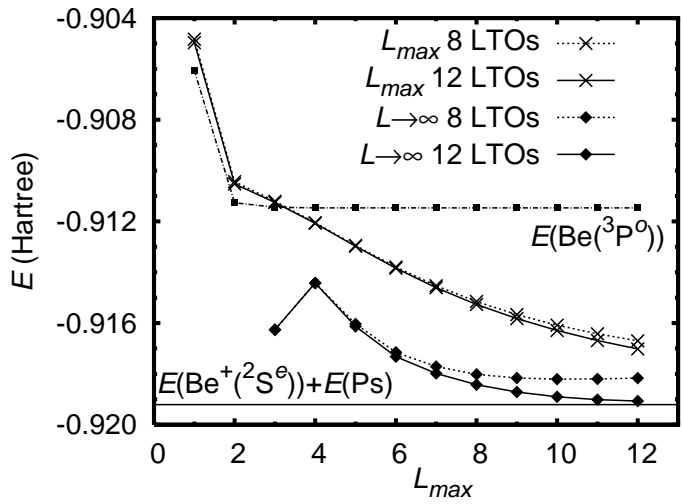}
}
\caption[]{ \label{fig:lmax}
The energy (in units of Hartree) of $e^+$-Be($^3$P$^o$) 
(crosses) and extrapolated energies (diamonds) as a function
of $L_{max}$ for two different calculations (with a minimum of
8 and 12 LTOs per $\ell$ respectively).
The Be($^3$P$^o$) two-electron energy is shown by the squares
for the 12 LTOs basis.  The threshold for binding
at $E$(Be$^+$)+$E$(Ps) is shown as solid line.
}
\end{figure}

There are two aspects where convergence can be incomplete.  Besides 
the size of $L_{\rm max}$, the number of LTOs for each value of $\ell$ 
still needs to be further increased.   This is best seen in
Figure~\ref{fig:lmax} which depicts the energies as a function of 
$L_{\rm max}$.  Two sets of $e^+$Be($^3$P$^o$) data are shown, the first 
shows the $e^+$Be($^3$P$^o$) energies from Table \ref{tab:belmax} which 
used 12 LTOs, and their extrapolated estimates using eq.~(\ref{extrap1}).   
The second set shows the results from initial exploratory
calculations also with $L_{\rm max} = 12$ and $L_{\rm int} = 3$, but only 
with a minimum of $8$ LTOs per $\ell$.  Even a cursory glance at 
Figure~\ref{fig:lmax} suggests that the CI prognosis for variationally
establishing the existence of a bound state is not promising.  Also shown
in Figure~\ref{fig:lmax} are the Be($^3$P$^o$) energies using the
underlying $e^+$Be($^3$P$^o$) basis without the positron (the 12 LTOs basis). 

Even though the valence annihilation rate from the largest calculation
is only $0.348 \times 10^9$ sec$^{-1}$, the slow convergence of $\Gamma_v$ 
with $L_{\rm max}$ suggests the presence of a well defined Ps cluster.
Fitting the four largest $\Gamma_v$ calculations in Table \ref{tab:belmax}
to the three-term form
\begin{equation}
\Delta \Gamma_L \approx \frac {A_{\Gamma}}{(L+{\scriptstyle \frac{1}{2}})^2} 
    + \frac {B_{\Gamma}}{(L+{\scriptstyle \frac{1}{2}})^3}
    + \frac {C_{\Gamma}}{(L+{\scriptstyle \frac{1}{2}})^4} \; ,  
\label{extrap2}
\end{equation}
results in an extrapolated annihilation rate of 
$\Gamma_v = 0.805 \times 10^9$ sec$^{-1}$.  However, this
value should be regarded as only notional.  After all,
if this system is indeed unbound, $\Gamma_v$ should approach
$\Gamma_{\rm Ps}=2.008 \times 10^9$ sec$^{-1}$
in the limit of an infinite radial basis \cite{mitroy05i}.

It is also noticed that $\langle r_p \rangle$ in Table \ref{tab:belmax}
steadily decreases as $L_{\rm max}$ increases.  This also should not be 
taken as an indicator of binding since the behavior of $\langle r_p \rangle$ 
with $L_{\rm max}$ is not straightforward.  For example, the PsH 
$\langle r_p \rangle$ decreases when $L_{\rm max}$ changes from 0 to 3, 
but then starts increasing as $L_{\rm max}$ increases from 3 to 9 
\cite{bromley02a,mitroy06a}.  Indeed, this is a trend noticed in CI 
calculations of positronic atoms with $IP<6.8$eV
(eg. $e^+$Li \cite{bromley02e}, $e^+$Ca and $e^+$Sr \cite{bromley02b,bromley06c}),
which have shown an initial $\langle r_p \rangle$
decrease to a minimum, which then increases as $L_{\rm max}\to\infty$.
The behavior of $\langle r_p \rangle$ with $L_{\rm max}$ for an unbound 
system is unknown.

\section{Conclusions}

A large-scale CI calculation of the $e^+$Be($^3$P$^o$) 
ground state has been performed.  The largest calculation
gave an energy that was $0.00219$ Hartree above the threshold
for binding.  Using an extrapolation method to estimate the
contribution from the higher-partial waves gave an energy that
was $0.00014$ Hartree above threshold.
By its very nature the present calculation is unable to give
definitive proof that the $e^+$Be($^3$P$^o$) system does not
have a bound state.  While a variational calculation of
the present kind can be used to give proof of binding, it
cannot be used to give proof of non-binding.  A converged 
calculation of low energy Be$^+$-Ps($1s$) scattering in the 
$^4$P$^o$ channel would be needed to establish lack of binding.  

What can be inferred from the present calculation is that a 
$e^+$Be($^3$P$^o$) bound state (assuming one existed) would 
have a structure similar to positronic lithium, i.e. it would 
consist of a Ps-like object weakly bound to the Be$^+$($2s$) core.  
The size of the CI expansion required to establish binding would 
be enormous.  For example, it is necessary to include 
orbitals with $L_{\rm max} \approx 30$ to get an energy that is lower 
than the Ps + Li$^+$ threshold \cite{bromley02e}.   
Another conclusion is that the existence of centrifugal barriers 
does seem to inhibit the binding of positrons to parent atoms 
with non-zero angular momentum.   

\section*{Acknowledgments}

The authors would like to thank Shane Caple of CDU for computer 
support.  Some preliminary calculations were performed on 
computing facilities made possible by the Research Corporation.
One author (MB) would like to thank Dr. Brett Esry for
discussions on three-body systems with non-zero angular momentum.
The authors would also like to thank Dr. Masanori Tachikawa for
access to unpublished data on positronic systems with non-zero angular momentum.



\begin{sidewaystable}
\centering
\caption[]{ \label{tab:belmax}
Results of CI calculations for $e^+$Be($^3$P$^o$) for a series of 
$L_{\rm max}$, with fixed $L_{\rm int}=3$ (the row denoted 12$^*$ 
was computed with $L_{\rm int}=2$).  The total number of electron and 
positron orbitals are identical and denoted by $N$.
The 3-body energy of the $e^+$-Be($^3$P$^o$) system, relative to
the energy of the Be$^{2+}$ core, is denoted by $E$ (in Hartree).
The threshold for binding is -0.9192086 Hartree, and
$\varepsilon$ gives binding energy (in Hartree) against
dissociation into Ps + Be$^+$($2s$).
The mean electron-nucleus distance $\langle r_e \rangle$,
the mean positron-nucleus distance $\langle r_p \rangle$,
and the mean electron-positron separation $\langle r^2_{ep} \rangle$
are given in units of a$_0$. The $\Gamma_v$ and $\Gamma_c$ columns
give the valence and core annihilation rates respectively (in units of $10^9$ sec$^{-1}$).
The $p_E$ column gives the power-series exponents from the 
$L_{\rm max},L_{\rm max}\!-\!1,L_{\rm max}\!-\!2$ energies.
The results in the row $\infty$ used eq.~(\ref{extrap1}) and 
eq.~(\ref{extrap2}) to estimate the $L_{\rm max} \rightarrow \infty$ 
correction.
}
\begin{tabular}{lcccccccccc} 
\hline \hline
$L_{\rm max}$&  $N$ &  $N_{CI}$ & $E$ & $\varepsilon$ & $\langle r_e \rangle$ &  $\langle r_p \rangle$ & $\langle r^2_{ep} \rangle$
                                                             & $\Gamma_c$  & $\Gamma_v$ & $p_E$ \\
\hline
  1 &  32 &  11325 & -0.9049669 & -0.0142417 & 2.70529 & 31.8236 & 1159.699 & 0.000020 & 0.000521 &   \\
  2 &  45 &  32580 & -0.9105271 & -0.0086815 & 2.70601 & 27.4961 & 920.4328 & 0.000087 & 0.004302 &  \\
  3 &  57 &  65694 & -0.9112555 & -0.0079531 & 2.72410 & 21.5600 & 619.5881 & 0.000284 & 0.020168 & 6.0406 \\
  4 &  69 & 108570 & -0.9120770 & -0.0071316 & 2.76099 & 16.3218 & 378.4203 & 0.000571 & 0.053933 & -0.479 \\
  5 &  81 & 153510 & -0.9129784 & -0.0062302 & 2.80833 & 13.2441 & 250.9955 & 0.000816 & 0.097421 & -0.462 \\
  6 &  93 & 201606 & -0.9138427 & -0.0053659 & 2.85800 & 11.6135 & 189.6602 & 0.000977 & 0.141967 & 0.2515 \\
  7 & 105 & 250350 & -0.9146093 & -0.0045993 & 2.90693 & 10.6898 & 157.1977 & 0.001076 & 0.184291 & 0.8378 \\
  8 & 117 & 300030 & -0.9152671 & -0.0039415 & 2.95389 & 10.1253 & 138.1528 & 0.001133 & 0.223404 & 1.2235 \\
  9 & 129 & 349710 & -0.9158238 & -0.0033848 & 2.99833 & 9.75778 & 125.9701 & 0.001166 & 0.259160 & 1.5001 \\
 10 & 141 & 399390 & -0.9162926 & -0.0029160 & 3.03992 & 9.50665 & 117.6386 & 0.001183 & 0.291698 & 1.7182 \\
 11 & 153 & 449070 & -0.9166867 & -0.0025219 & 3.07841 & 9.32966 & 111.6870 & 0.001191 & 0.321221 & 1.9055 \\
 12 & 165 & 498750 & -0.9170182 & -0.0021904 & 3.11361 & 9.20197 & 107.2937 & 0.001193 & 0.347951 & 2.0775 \\
12$^{*}$ & 165 & 334248 & -0.9169903 & -0.0022183 & 3.11398 & 9.20391 & 107.3440 & 0.001192 & 0.347880 & \\
\hline  
 $\infty$ &  &     & -0.9190697 & -0.0001390 & &  &   &                                     & 0.805027  & \\
\hline \hline
\end{tabular}
\end{sidewaystable}

\end{document}